\documentclass[preprint,showpacs,preprintnumbers,amsmath,amssymb]{revtex4}

\usepackage{graphicx}
\usepackage{dcolumn}
\usepackage{bm}
\usepackage{textcomp}

\begin{document}

\preprint{APS/123-QED}

\title{Sol-Gel Derived Ferroelectric Nanoparticles \\Investigated by Piezoresponse Force Microscopy}

\author{F. Johann$^1$}
\author{T. Jungk$^1$}
\author{S. Lisinski$^2$}
\author{\'{A}. Hoffmann$^1$}
\author{L. Ratke$^2$}
\author{E. Soergel$^1$}
\email{soergel@uni-bonn.de}

\affiliation{$^1$Institute of Physics, University of Bonn,
Wegelerstra\ss e 8, 53115 Bonn, Germany}

\affiliation{$^2$Institute for Material Physics in Space,
DLR, Linder H\"{o}he, 51147 Cologne, Germany}

\date{\today}

\begin{abstract}
Piezoresponse force microscopy (PFM) was used to investigate the ferroelectric properties of sol-gel derived LiNbO$_3$ nanoparticles. To determine the degree of ferroelectricity we took large-area images and performed statistical image-analysis. The ferroelectric behavior of single nanoparticles was verified by poling experiments using the PFM tip. Finally we carried out simultaneous measurements of the in-plane and the out-of-plane piezoresponse of the nanoparticles, followed by measurements of the same area after rotation of the sample by 90$^{\circ}$ and 180$^{\circ}$. Such measurements basically allow to determine the direction of polarization of every single particle.
\end{abstract}


\maketitle


In the past decade nonlinear-optical materials have gained great interest due to their potential for applications such as frequency converters via second-harmonic generation~\cite{Fej92} or difference frequency mixing~\cite{Lim91}. As a matter of course the boom of nanoparticle research has also reached this material class. Consequently, nanoparticles of lithium niobate (LiNbO$_3$) and lithium tantalate (LiTaO$_3$), both materials possessing high nonlinear-optical coefficients~\cite{Sho97}, are currently investigated~\cite{Tru07,Bau04,Ski04}. The use of the sol-gel process for the fabrication of nanoparticles of the desired dimensions has additionally boost the research in this field. Although several techniques for the characterization of nanoparticles are available, such as electron microscopy~\cite{Gra97,Uts03}, X-ray photoelectron spectroscopy~\cite{Lop06}, X-ray diffraction~\cite{Som07} or Raman spectroscopy~\cite{Dob02}, they can not provide information about their ferroelectricity. The latter, however, is a mandatory property for the above mentioned optical applications. Lately, piezoresponse force microscopy (PFM) has become a standard technique for the investigation of ferroelectric domains on the nanometer scale~\cite{Alexe,Jun06,Jun07,Jun08}. Since PFM is not limited to specific crystallographic orientations~\cite{Jun09}, it is ideally suited for the investigation of randomly oriented ferroelectric nanoparticles~\cite{Gau06}.


To identify the orientation of a ferroelectric particle by PFM it is necessary to determine three independent spatial directions of its piezoresponse. PFM allows to map out-of-plane and in-plane piezoresponse simultaneously as deflection and torsion of the cantilever. Thus, recording two such image pairs of the same area at different angles between cantilever and sample yields the required information. Unfortunately, in-plane piezoresponse leads also to buckling of the cantilever which, same as deflection, is recorded as a vertical signal~\cite{Jun06a}. A discrimination of those two movements and hence a clear assignment of the driving forces recorded in the vertical signal is therefore mandatory.

Vector piezoresponse force microscopy~\cite{Kal06} does not provide full 3D information on the polarization distribution. Since this technique is based on one image pair only acquired within a single scan, only one component of the in-plane driving forces is mapped. Furthermore the vertical signal is not split up into buckling and deflection.
Few studies showing PFM images differenced by a 90$^{\circ}$ rotation of the sample have been reported~\cite{Eng99,Guy09}, however, deflection was never discriminated from buckling so far. Thus, an unambiguous attribution of the measured signals to the effective piezoresponse was not possible.

In this contribution we present investigations on randomly oriented sol-gel derived $\rm LiNbO_3$ nanoparticles using PFM. The primary goal was to determine the degree of ferroelectricity of the nanoparticles fabricated by this method. Even more, we were able to
discriminate deflection from buckling in the vertical signal by controlled rotation of the sample with respect to the cantilever and appropriate image processing. This result signifies a real progress for the PFM technique itself.


For the PFM measurements we used a commercial scanning force microscope (SMENA from NT-MDT) upgraded with a high-precision rotation stage~\cite{Jun09}. By careful adjustment of the rotation axis with respect to the position of the tip a shift of the image by only few microns upon rotation by 180$^{\circ}$ can be achieved. We could thereby record the same area at different angles between cantilever and sample.

The samples under investigation were LiNbO$_3$-xerogels prepared by the sol-gel method~\cite{Lis06}. For their synthesis following components were used: lithiumethoxid and niobiumethoxid as precursors, acetic acid as catalyst, and  ethanol as solvent. For stochiometric xerogels the molar ratio between lithiumethoxid and niobiumethoxid was 1:1 (for congruent ones 1:1.9). The synthesized materials were sintered at 700$^{\circ}$C for 48\,hours to obtain crystalline LiNbO$_3$. In case of congruent LiNbO$_3$-xerogels a second phase (non-ferroelectric LiNb$_3$O$_8$) emerges during the heat treatment. We investigated both, stoichiometric and congruent samples.

Figure~\ref{fig:johann01} shows simultaneously recorded topography and vertical PFM signal of both samples. The topographic images reflect the expected humped structure, as it was also seen in scanning electron microscopy imaging~\cite{Lis09}, and is similar for both samples (Figs~\ref{fig:johann01}(a,d)). The PFM images (Figs~\ref{fig:johann01}(b,e)), however, have a very different appearance: the nanoparticles seem to be transferred to flatland~\cite{flatland}. This is because the piezoresponse of every nanoparticle is same across the whole particle, leading to one grey level according to its orientation. Obviously, in Fig.~\ref{fig:johann01}(b) nearly every nanoparticle shows a different piezoresponse, whereas in Fig.~\ref{fig:johann01}(e) many particles show the same mean grey value indicating their being non-ferroelectric.

In the next step we performed a statistical analysis of the PFM images. We therefore readout the grey level of every pixel ($512 \times 512$) of the images and assigned them a value between $-1$ (black) and $+1$ (white). For the purely ferroelectric sample the analysis yields grey levels in a homogeneously distribution (Fig.~\ref{fig:johann01}(c)) as expected from randomly oriented particles~\cite{Har02}. For the congruent sample, containing also non-ferroelectric particles the statistics results in a very different grey level distribution (Fig.~\ref{fig:johann01}(f)). Here, the amount of non-ferroelectric particles is approx.\ 80\% which can be estimated from the height of the central peak at position 0. Note that the substructure in the distribution is due to missing statistics, the number of particles imaged being too small.

\begin{figure}
\includegraphics{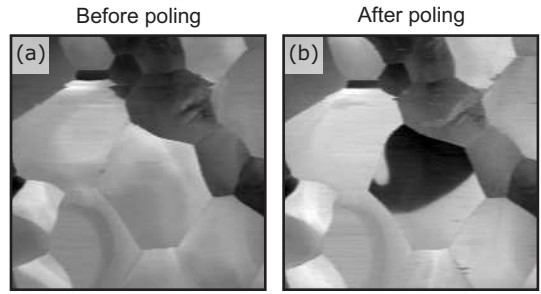}
\caption{Local poling of a LiNbO$_3$ nanoparticle: (a) shows the initial state and (b) the domain configuration after partial poling of a nanoparticle. PFM image size is $6 \times 6$\,\textmu m$^2$ each.}
\label{fig:johann02}
\end{figure}

A proof for the particles' ferroelectricity is their poling behavior. For local poling with the PFM tip we chose a particle whose orientation was out of plane, i.\,e., a particle showing up bright in the vertical image. We applied a positive voltage to the tip (+100\,V for 60 seconds) to partially reverse the orientation of the particle. Figure~\ref{fig:johann02} shows the result of such a poling experiment; (a): initial state and (b): after poling. Two observations can be made: (i) the particle has not been completely poled and (ii) poling is confined within the limitation of the particle. Whereas complete poling could be achieved by increasing voltage or writing time, we never observed a neighboring particle to be poled on contrary to results from oriented PZT thin films~\cite{Shi00}. Indeed, the poling behavior of adjoining particles is of great importance for applications such as data storage devices.

Finally, we want to emphasize the potential of the high-precision rotation stage for determining the polarization of single nanoparticles. We therefore took PFM images at three angles between sample and cantilever: at 0$^{\circ}$, 90$^{\circ}$ and 180$^{\circ}$. The vertical (V) and the lateral (L) signals were recorded simultaneously, thus resulting all together in six PFM images. The in-plane components of the piezoresponse can be directly obtained from two lateral images unambiguously correlated to the torsion (T) of the cantilever: T(0$^{\circ})$ and T(90$^{\circ})$. The out-of-plane component can only be obtained after splitting the vertical signal into deflection (D) and buckling (B). We therefore used two images rotated by $180^{\circ}$: $\rm D(0^{\circ}) =  \frac{1}{2}\{V(0^{\circ}) + V(180^{\circ})\}$. The ''$+$'' is because buckling experiences a phase shift by $\pi$ upon rotation by 180$^{\circ}$. In the same manner buckling can be extracted from the vertical signal: $\rm B(0^{\circ}) =  \frac{1}{2}\{V(0^{\circ}) - V(180^{\circ})\}$. The requirement that $\rm T(90^{\circ}) = B(0^{\circ})$ finally allows to verify the image processing.

\begin{figure}
\includegraphics{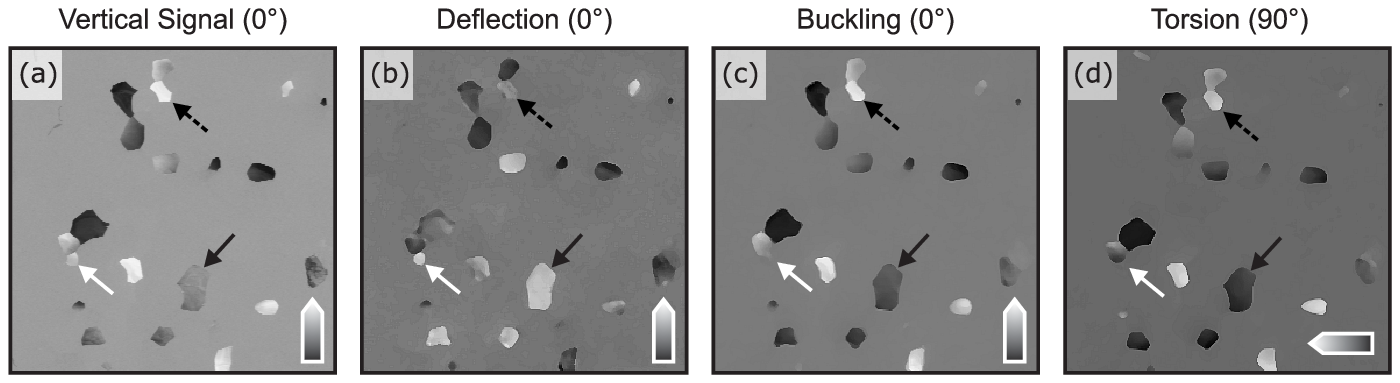}
\caption{Vertical PFM signal (a) and the discriminated deflection (b) and buckling (c). (d) shows the torsion recorded at $90^{\circ}$. The three arrows (white, black and dashed black) indicate selected particles analyzed in detail. The direction of the cantilever is indicated at the bottom on the right. Image size is $20 \times 20$\,\textmu m$^2$ each.}
\label{fig:johann03}

\end{figure}

Figure~\ref{fig:johann03} shows results on the congruent sample as  measured ((a) and (d)) and image processed ((b) and (c)) data.  Obviously Figs~\ref{fig:johann03}(c) and (d) exhibit the very same in-plane piezoresponse ($\rm T(90^{\circ}) = B(0^{\circ})$), hence image processing for the discrimination of deflection and buckling from the vertical readout channel works reliably. In the following we will analyze exemplarily three selected particles to demonstrate the possibilities of this recording/calculation scheme:
\begin{itemize}
\setlength{\itemsep}{0cm}
  \item[\textbf{-}] There is one small particle (indicated by the white arrow) whose polarization direction is oriented almost perfectly out of plane. This can be deduced from the fact that this particle is seen in the deflection image D($0^\circ$), but not in the lateral images T($90^\circ$)~(d) and T($0^\circ$) (not shown).
  \item[\textbf{-}] The particle indicated by the black arrow is characterized by its faint contrast in the vertical signal, but turns out to show a distinct contrast in both, D($0^{\circ}$) as a bright and B($0^{\circ}$) as a dark particle. Obviously, the orientation of the particle is such that the contributions from deflection and buckling cancel mostly out each other, therefore nearly no contrast is seen in the vertical image.
  \item[\textbf{-}] The dashed black arrow finally points towards a particle whose direction of polarization lies along the axis of the cantilever, as it shows a contrast in B($0^{\circ}$) but is nearly invisible in D($0^{\circ}$).
\end{itemize}

The crucial question now is: can one unambiguously attribute the exact orientation to every particle using PFM? A simple answer is not quite possible as exposed in the following. Decidedly, one can determine direction and magnitude of the piezoresponse of every nanoparticle using a rotation facility together with a careful calibration of the microscope. However, to attribute this piezoresponse to the orientation of the particle is not straight forward. We want to give two reasons for this comment: i) all materials has several piezoelectric tensor elements, and the PFM signals detected originate from their interplay, which strongly impedes the interpretation of the measurements;  ii) the electric field generated by the tip is not only directed along the tip axis but has components in all spatial directions. Since every component can lead to a piezoresponse of the sample~\cite{Jun09} an analysis of the obtained PFM signals becomes further more complicated. The unambiguous determination of the orientation by PFM is therefore at least very challenging, and might not be accomplished without complex numerical methods.

In conclusion, we have accomplished a detailed analysis of sol-gel derived ferroelectric lithium niobate nanoparticles. By comparative measurements using samples containing a large amount of ballast particles, we could ensure that nanoparticles fabricated by sol-gel technique method are ferroelectric. Furthermore, we used those samples to demonstrate the advanced possibilities of a high-precision rotation stage for the interpretation of PFM images including the discrimination of deflection and buckling, a mandatory requirement to allow for the determination of the direction of polarization.


{\small {\bfseries Acknowledgments} Financial support from the Deutsche Telekom AG is gratefully acknowledged.}


\newpage

\begin{figure}
\includegraphics{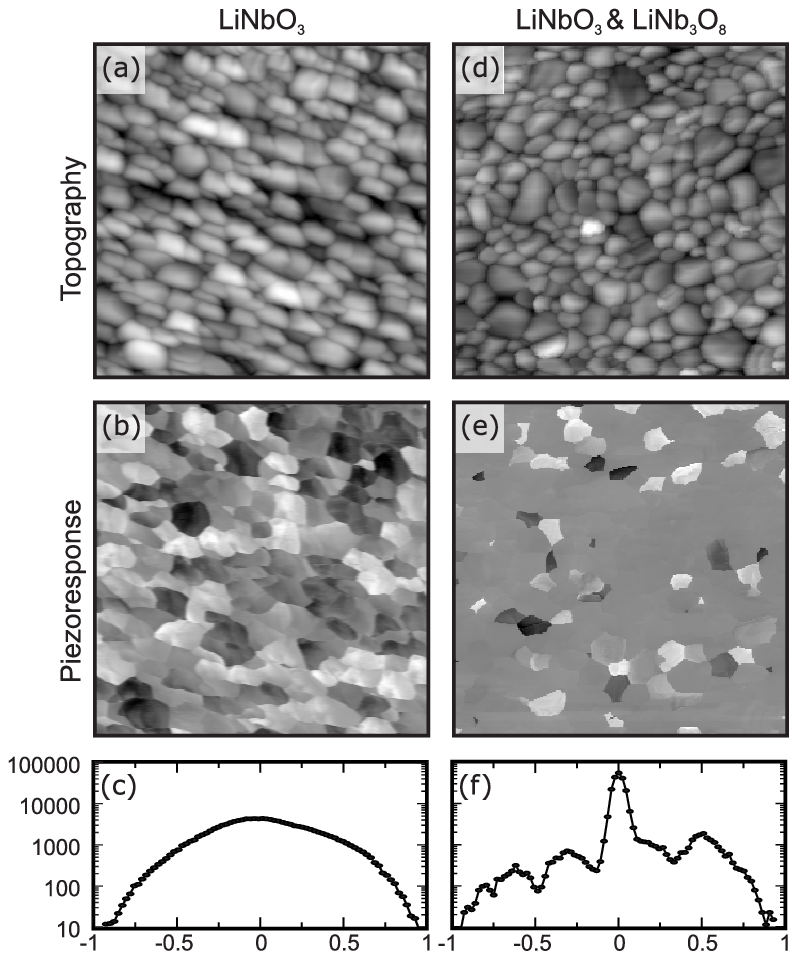}
\caption{Comparative measurements of the topography (a, d) and the piezoresponse (b, e) of two LiNbO$_3$ nanoparticle samples. Whereas (a) and (b) were recorded on a sample containing only ferroelectric particles, the images in (d) and (e) were taken using a sample with a large amount of ballast particles. The graphs in (c) and (f) show the grey level statistics of the piezoresponse images (b) and (e) respectively. Image size is $20 \times 20$\,\textmu m$^2$ each.}
\end{figure}
\label{fig:johann01}

\clearpage

\clearpage

\begin{thebibliography}{}

\bibitem{Fej92}
M.\,M.\,Fejer, G.\,A.\,Magel, D.\,H.\,Jundt, and R.\,L.\,Byer, IEEE J.\
Quantum Elect.\ \textbf{28}, 2631 (1992).

\bibitem{Lim91}
E.\,J.\,Lim, H.\,M.\,Herka, M.\,L.\,Bortz, and M.\,M.\,Fejer,
Appl.\ Phys.\ Lett.\ \textbf{59}, 2207 (1991).

\bibitem{Sho97}
I.\,Shoji, T.\,Kondo, A.\,Kitamoto, M.\,Shirane, and R.\,Ito,
J.\ Opt.\ Soc.\ Am.\ B \textbf{14}, 2268 (1997).

\bibitem{Tru07}
J.\,Trull, C.\,Cojocaru, R.\,Fischer, S.\,M.\,Saltiel, K.\,Staliunas,  R.\,Herrero,  R.\,Vilaseca, D.\,N.\,Neshev, W.\,Krolikowski, and Y.\,S.\,Kivshar, Opt.\ Express \textbf{24}, 15868 (2007).

\bibitem{Bau04}
M.\,Baudrier-Raybaut, R.\,Haidar, Ph.\,Kupecek, Ph.\,Lemasson, and E\,~Rosencher,
Nature \textbf{432}, 374 (2004).

\bibitem{Ski04}
S.\,E.\,Skipetrov, Nature \textbf{432}, 285 (2004).

\bibitem{Gra97}
K.\,C.\,Grabar, K.\,R.\,Brown, C.\,D.\,Keating, S.\,J.\,Stranick, S.-L.\,Tang, and M.\,J.\,Natan,
Anal.\ Chem.\ \textbf{69}, 471 (1997).

\bibitem{Uts03}
S.\,Utsunomiya and R.\,C.\,Ewing,
Environ.\ Sci.\ Technol.\ \textbf{37}, 786 (2003).

\bibitem{Lop06}
I.\,Lopez-Salido, D.\,C.\,Lim, R.\,Dietsche, N.\,Bertram, and Y.\,D\,Kim,
J.\ Phys.\ Chem.\ B \textbf{110}, 1128 (2006).

\bibitem{Som07}
V.\,Somani and S.\,J.\,Kalita,
J.\ Electroceram.\ \textbf{18}, 57 (2007).

\bibitem{Dob02}
P.\,S.\,Dobal and R.\,S.\,Katiyar,
J.\ Raman Spectrosc.\ \textbf{33}, 405 (2002).

\bibitem{Alexe}
M.\,Alexe and A.\,Gruverman, eds.,
{\it Nanoscale Characterisation of Ferroelectric Materials} (Springer, Berlin; New York, 2004) 1st ed.

\bibitem{Jun06}
T.\,Jungk, \'{A}.\,Hoffmann, and E.\,Soergel,
Appl.\ Phys.\  Lett.\ \textbf{89}, 163507 (2006).

\bibitem{Jun07}
T.\,Jungk, \'{A}.\,Hoffmann, and E.\,Soergel,
J.\ Microsc.\ \textbf{227}, 72 (2007).

\bibitem{Jun08}
T.\,Jungk, \'{A}.\,Hoffmann, and E.\,Soergel,
New  J.\ Phys.\ \textbf{10}, 013019 (2008).

\bibitem{Jun09}
T.\,Jungk, \'{A}.\,Hoffmann, and E.\,Soergel,
New  J.\ Phys.\ \textbf{11}, 033029 (2009).

\bibitem{Gau06}
B.\,Gautier and V.\,Bornand,
Thin Solid Films \textbf{515}, 1592 (2006).

\bibitem{Jun06a}
T.\,Jungk, \'{A}.\,Hoffmann, and E.\,Soergel,
Appl.\ Phys.\  Lett.\ \textbf{89}, 042901 (2006).

\bibitem{Kal06}
S.\,V.\,Kalinin, B.\,J.\,Rodriguez, S.\,Jesse, J.\,Shin, A.\,P.\,Baddorf,
P.\,Gupta, H.\,Jain, D.\,B.\,Williams, and A.\,Gruverman, Microscopy
and Microanalysis \textbf{12}, 206 (2006).

\bibitem{Eng99}
L.\,M.\,Eng, H.-J.\,G\"{u}ntherodt, G.\,A.\,Schneider, U.\,K\"{o}pke, and J.\,M.\,Salda\~{n}a,
Appl.\ Phys.\ Lett.\ \textbf{74}, 233 (1999).

\bibitem{Guy09}
J.\,Guyonnet, H.\,B\'{e}a,  F. Guy, S.\,Gariglio,  S. Fusil,  K. Bouzehouane, J.-M.\,Triscone, P.\,Paruch
Appl. Phys. Lett. \textbf{95}, 132902 (2009).

\bibitem{Lis06}
S.\,Lisinski, D.\,Schaniel, L.\,Ratke, and Th.\,Woike,
Chem.\ Mater.\ \textbf{18}, 1534 (2006).

\bibitem{Lis09} S.\,Lisinski, PhD thesis ''Synthese und Charakterisierung von ferroelektrischen Xerogelen'' RWTH Aachen (2008)

\bibitem{flatland}
Edwin A.\,Abbott ''Flatland: A Romance of Many Dimensions'', (1884).


\bibitem{Har02}
C. Harnagea, A. Pignolet, M. Alexe, and D. Hesse,
Integrated Ferroelectrics \textbf{44}, 113 (2002)

\bibitem{Shi00}
H.\,Shin, J.\,Woo, S.\,Hong, J.\,U.\,Jeon, Y.\,E.\,Pak, K.\,No,
Integrated Ferroelectrics \textbf{31}, 163 (2000).



\end{thebibliography}
\end{document}